\documentclass[conference]{IEEEtran}
\IEEEoverridecommandlockouts
\usepackage[utf8]{inputenc}

\usepackage{balance,multicol}
\usepackage{color}
\usepackage{pdfsync}
\usepackage{ifthen}
\usepackage[ruled,vlined]{algorithm2e}
\usepackage{svg}
\usepackage{cool}
\usepackage{makeidx}
\usepackage{verbatim}
\usepackage{stfloats}
\usepackage{algorithmic}
\usepackage{amsmath}

\usepackage{amsfonts} 
\usepackage{multirow,array} 
\usepackage{booktabs}
\usepackage{tikz}
\usepackage{enumitem}
\usetikzlibrary{automata,arrows,positioning,calc}
\usepackage{lipsum}
\usepackage{threeparttable} 
\usepackage{blindtext}
\PassOptionsToPackage{hyphens}{url}
\usepackage[bookmarks=false]{hyperref}
\usepackage{gensymb}
\usepackage{chemformula}
\usepackage{tabu} 
\usepackage{tabularx}
\usepackage{makecell}
\usepackage{cite}
\usepackage{siunitx}
\usepackage{colortbl}
\usepackage{booktabs}
\usepackage{threeparttable}

\usepackage{eso-pic}

\usepackage{cite}
\usepackage{amsmath,amssymb,amsfonts}
\usepackage{algorithmic}
\usepackage{graphicx}
\usepackage{textcomp}
\usepackage{xcolor}
\def\BibTeX{{\rm B\kern-.05em{\sc i\kern-.025em b}\kern-.08em
    T\kern-.1667em\lower.7ex\hbox{E}\kern-.125emX}}
\begin{document}
\title{Dynamic Edge Server Selection in Time-Varying Environments:\\
A Reliability-Aware Predictive Approach}

\author{
\IEEEauthorblockN{
    Jaime Sebastian Burbano,
    Arnova Abdullah,
    Eldiyar Zhantileuov,
    Mohan Liyanage,
    Rolf Schuster
}
\IEEEauthorblockA{ University of Applied Sciences and Arts, Dortmund, Germany\\
Email: \{jaime.burbanovillavicencio, arnova.abdullah, eldiyar.zhantileuov, mohan.liyanage, rolf.schuster\}@fh-dortmund.de}
}

\makeatletter
\AddToShipoutPicture*{%
  \put(55,770){%
    \parbox[t]{\textwidth}{%
      \raggedright
      \fontsize{8}{10}\selectfont
      Accepted for presentation at the 7th International Conference on Advancements in Computing (ICAC 2025), Colombo, Sri Lanka. %
    }%
  }%
}
\makeatother

\maketitle

% ---- IEEE copyright notice (visible and left-aligned) ----

\IEEEpubid{\begin{minipage}{\columnwidth}
    \vspace{10pt}
    \hspace{-4.7cm}
    \footnotesize 979-8-3315-6222-9/25/\$31.00~\copyright~2025 IEEE
\end{minipage}}
\IEEEpubidadjcol

\begin{abstract}
Latency-sensitive embedded applications increasingly rely on edge computing, yet dynamic network congestion in multi-server architectures challenge proper edge server selection. This paper proposes a lightweight server-selection method for edge applications that fuses latency prediction with adaptive reliability and hysteresis-based handover. Using passive measurements (arrival rate, utilization, payload size) and an exponentially modulated rational delay model, the proposed Moderate Handover (MO-HAN) method computes a score that balances predicted latency and reliability to ensure handovers occur only when the expected gain is meaningful and maintain reduced end-to-end latency. Results show that MO-HAN consistently outperforms static and fair-distribution baselines by lowering mean and tail latencies, while reducing handovers by nearly 50\% compared to pure opportunistic selection. These gains arise without intrusive instrumentation or heavy learning infrastructure, making MO-HAN practical for resource-constrained embedded devices.
\end{abstract}


\section{Introduction}
Modern embedded IoT systems are increasingly tasked with computation-intensive and latency-sensitive workloads, such as collaborative robotic manipulators for industrial automation, high-resolution inspection pipelines, mixed/augmented reality overlays, and perception modules in autonomous mobile robots and UAVs. These workloads often impose strict requirements, as even occasional excessive delay can
degrade control stability, impair perception quality, or diminish user experience. Despite continuous advances in embedded processors, the performance gap between application demands and local compute capacity is widening, making exclusive reliance on onboard execution infeasible.
 
Offloading task execution to nearby edge servers has thus emerged as a natural design choice. Edge computing provides low-latency access to high-performance resources while preserving data locality and reducing reliance on distant cloud infrastructures. However, the assumed determinism of edge resources is misleading. In multi-tenant deployments, edge servers and their interconnecting networks are subject to dynamic interference caused by temporarily increased traffic, bursty request arrivals, container scheduling delays, background workloads and queue build-ups~\cite{bittencourt2017mobility}. These factors introduce latency variance and heavy-tailed outliers, which can trigger high delays for embedded devices relying on timely responses, ultimately degrading system performance.

 \begin{figure}[htbp]
\centering
 \includegraphics[width=0.9\linewidth]{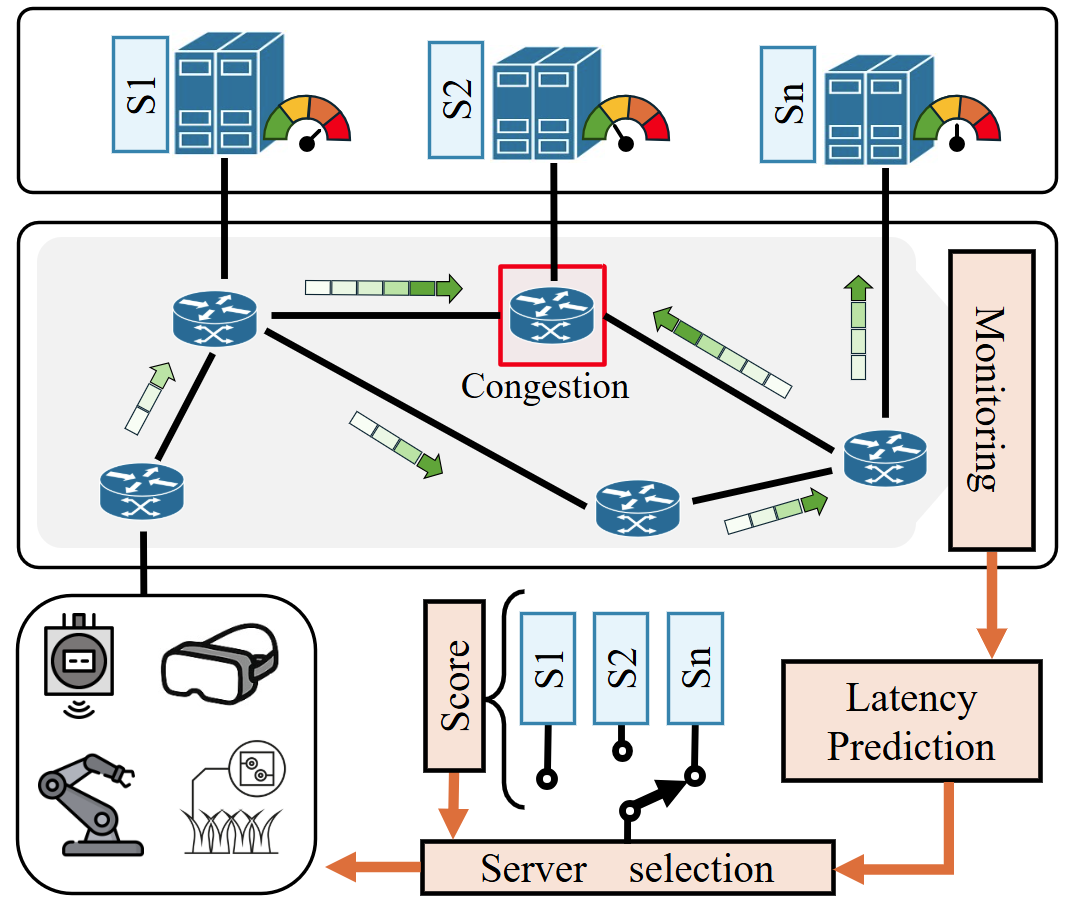}
\caption{Block diagram of the proposed server selection methodology. In time-varying conditions, sporadic congestion in network paths to a certain edge server leads to elevated latency.}
\label{fig_intro}
\end{figure}
 
When multiple edge servers coexist within a network, server selection becomes a moving target. Static policies (e.g., nearest edge) and purely reactive choices are weak under time-varying conditions; they either fail to anticipate emerging congestion or oscillate between servers, further destabilizing edge performance and posing overhead on embedded devices. Specifically, continuous handover between edge servers (e.g., when an embedded device connects to a different edge server) results in higher energy costs, extra latency delays, and overall reduced network performance ~\cite{li2019delay}. This motivates the need for a lightweight, prediction-driven, and reliability-aware server selection mechanism that anticipates communication latency to each available edge, but also accounts for historical deviations by reducing trust in servers that repeatedly fail to deliver the predicted response times, while minimizing unnecessary handovers.

This paper proposes a methodology that combines network monitoring, latency prediction, and adaptive reliability tracking into a unified server selection workflow. Non-intrusive link measurements are used to estimate the latency to candidate servers, while reliability scores are dynamically updated based on prediction violations. A hysteresis-based decision policy then integrates these metrics to trigger server handovers only when they yield significant performance improvements. Figure~\ref{fig_intro} illustrates the edge architecture under time-varying network conditions and the workflow of the proposed methodology.
 
The specific contributions of this paper include:

\begin{itemize}
    \item \textbf{Prediction-Guided Server Selection:}  
    We formulate edge server selection as a prediction-guided decision problem in embedded devices under a time-varying network and compute conditions, focusing on latency.

    \item \textbf{Adaptive Reliability Modeling:}  
    We introduce an online reliability metric, updated via exponential moving average, that penalizes recurrent prediction violations and captures dynamic edge behavior.

    \item \textbf{Switching Awareness with Hysteresis:}  
    We design a low-overhead selection policy with threshold-based switching, ensuring handover occurs only if the expected performance is meaningfully better. This prevents oscillations and reduces overheads, making the policy suitable for resource-constrained devices.
\end{itemize}

\section{Related work} \label{sec:relwo}
A central challenge in edge computing lies in selecting the most suitable server under dynamic conditions, while avoiding excessive handovers that degrade performance and increase overhead. Prior research on server selection has explored a wide spectrum of strategies, ranging from analytical heuristics to advanced machine learning–based approaches. In what follows, we provide a structured review of the most relevant contributions in this area, highlighting their strengths, limitations, and applicability to real-time embedded scenarios.

%One of the main challenges of edge computing applications is choosing the best server without frequently moving between the available edge servers. Previous studies on various server selection techniques have examined a variety of strategies, from analytical to machine learning-based approaches. This section provides a thorough analysis of the most relevant studies in the area.
%This section provides a thorough analysis of various server selection methods and the server switching issue. Additionally, prior research on task offloading for embedded edge computing applications is reviewed.

%\subsection{Server Selection Approaches}

Analytical and heuristic models provide foundational approaches for edge server selection by evaluating performance metrics such as latency, energy consumption, and resource utilization. Zhang et al.\cite{zhang2021novel} proposed a heuristic Genetic Algorithm and Simulated Annealing–based
method for edge Server Selection (GASS) based on consumed energy and delay. While this method outperforms traditional heuristic algorithms, it presents high complexity. A MEC orchestrator was developed in \cite{9936916} for 5G applications, optimizing server selection based on server resource availability and the communication channel state. 

Several studies have explored server selection for IoT applications, targeting delay minimization or cost reduction \cite{10943239,seeme,10710290,s25113443}. Imanaka et al.~\cite{10943239} proposed a preventive start-time optimization model to minimize delay with fault tolerance, while Dou et al.~\cite{seeme} introduced a location-aware approach for low-latency computation. Comparative analyses further show that Least Remaining CPU Cycles (LRC) and Fewest Active Tasks (FAT) outperform baseline methods under varying workloads~\cite{10710290,s25113443}.

In addition to analytical and heuristic approaches, machine learning (ML), reinforcement learning (RL), and federated learning (FL) techniques have been applied to server selection. For instance, in vehicular edge computing, TS-LSTM predicts vehicle trajectories while a Deep Q-Network (DQN) selects target servers to ensure service continuity \cite{trajectory}. TRUST-ME \cite{fi16080278} employs a dual trust-based RL framework for fair and reliable server selection in IoT scenarios, and quadruple Q-learning has been applied to minimize latency variance in latency-sensitive applications such as cloud gaming \cite{fair}. Kim et al. \cite{joint-edge} proposed a federated learning framework for joint MEC server selection and dataset management, optimizing accuracy, latency, and energy efficiency in mobile traffic prediction.

Several studies have explored Deep Reinforcement Learning (DRL) for edge server selection. These works include Markov Decision Process (MDP) based server selection with the lowest overall cost \cite{liu2021deep}, TD3-based frameworks for 5G-enabled industrial and C-ITS applications \cite{saad2025twin}, %multi-objective DRL for edge-cloud server optimization \cite{opti}, 
and adaptive DRL combined with meta-heuristic optimization for dynamic task offloading \cite{vidya2025dynamic}. These approaches collectively demonstrate the effectiveness of RL in managing latency, load balancing, and resource utilization across heterogeneous edge computing networks. However, these DRL-based approaches largely overlook integrating explicit latency prediction and minimizing frequent server switching.
%Additionally, they do not fully address heterogeneous dynamic workloads in realistic edge computing environments.

In edge computing, frequent server switching can increase both latency and energy consumption. To address this, \cite{8761232} formulated the MinEn problem and proposed a dynamic server switching algorithm that leverages device distribution prediction. While primarily focused on energy efficiency, this method also reduces the frequency of server switching. 

Recently, strategies that are aware of switching have been proposed to enhance stability in systems. Zhao et al.~\cite{zhao2024cloud} introduced a cloud-edge cooperative Distributed Model Predictive Control (DMPC) framework for vehicle platoons, which effectively reduces unnecessary switching between servers. Ortiz \cite{ortiz2025learning} addresses server switching in MEC using RL with learning and forgetting mechanisms, allowing mobile units to adaptively select servers in real time while discarding outdated information, thus maintaining low latency and energy consumption under dynamic conditions. %However, current research has not fully addressed the issue of stability in edge server selection when faced with dynamic workloads. 
Implementing prediction-based selectors with hysteresis could further stabilize server selection and decrease frequent switching under these conditions.

\section{System Model}

\subsection{Network Monitoring and Latency Prediction}

Let an embedded device $\mathbf{D}$ be connected to an edge infrastructure $\mathbf{E}$ composed of candidate edge servers $\{\phi_1,\phi_2,\dots,\phi_J\}$. At dynamic request time $t$, $\mathbf{D}$ generates a request of size $\gamma$ bytes to be transmitted and processed by one of the servers $\phi_j \in \mathbf{E}$. The path between $\mathbf{D}$ and $\phi_j$ traverses $n_j$ intermediate routers, each characterized by its instantaneous network utilization $\Phi_{k}$ and arrival rate $\Upsilon_{k}$, where $k=1,\dots,n_j$. These variables capture the congestion state and traffic intensity of each hop in the path.

To estimate communication latency, $\mathbf{D}$ (or its monitoring proxy) collects measurements of traffic characteristics, such as packet arrival rates and link utilization, using passive monitoring techniques (e.g., flow statistics, TCP acknowledgment timing, or in-network observability agents)~\cite{Taherizadeh2017,Caiazza2020}. Such measurements form a lightweight feature vector $\mathbf{x}$, which is periodically updated during system operation.

End-to-end latency in edge networks results from nonlinear interactions among traffic rate, link utilization, and payload size. Classical queuing and regression models often fail to capture abrupt latency escalations near saturation. To address this, we proposed a rational delay model~\cite{liyanage2025} with an exponential modulation term that reflects multiplicative delay growth under heavy load. The model evaluation and performance metrics are detailed in our previous work~\cite{liyanage2025}. In this article, we extend that approach by integrating latency prediction with adaptive reliability scoring and hysteresis-based switching, ensuring a balance between short-term responsiveness and long-term stability.

% The end-to-end latency distribution in edge networks is shaped by nonlinear interactions between traffic arrival rates, link utilization, and payload size. Classical queuing-theoretic or linear regression models often fail to capture compounding effects such as sudden latency spikes under near-saturation conditions. To this end, we have previously extended the \emph{rational delay model} with an exponential modulation term that captures multiplicative escalation under heavy traffic [undisclosed].

The predicted one-hop latency $\hat{y}$ is defined as:
\begin{equation}
\hat{y}(\mathbf{x}) = 
\left(
\frac{\sum_{i=1}^{m} a_i x_i}{1 + \sum_{j=1}^{n} b_j x_j + c}
\right)
\cdot \exp(d \cdot x_3),
\label{eq:rational_exp_delay_model}
\end{equation}

where:
\begin{itemize}
    \item \( m = n \) denotes the number of monitored parameters,
    \item \( x_{i,j} \in \{\gamma , \Phi, \Upsilon \}\), 
    %= request size in bytes,
    %\item \( x_2 \) = instantaneous link utilization (\%),
    %\item \( x_3 \) = packet arrival rate (packets/s),
    \item \( a_i, b_j, c,~ and ~d \) are model coefficients estimated from training data.
\end{itemize}

The predicted end-to-end latency from device $\mathbf{D}$ to server $\phi_j$ is obtained by summing per-hop predictions:
\begin{equation}
\hat{T}_{D \rightarrow \phi_j} = 
\sum_{k=1}^{n_j} \hat{y}(\mathbf{x}_k),
\label{eq:e2e_model}
\end{equation}

\subsection{Adaptive Reliability and Moderated Handover Modeling}
\label{sec:reliability_modeling}

While latency prediction is essential, real-world edge environments demand robustness to model inaccuracy, degraded performance, or unresponsiveness. To address this, we propose to augment the server selection logic by incorporating dynamic reliability scores for each candidate edge server and a switching threshold with hysteresis to reduce unnecessary edge server handovers.

Each edge server \( \phi{_j} \in \{\mathbf{E}\} \) is assigned an initial reliability score $r_j$. This value reflects prior expectations about the edge server’s stability or performance and may be configured manually or based on empirical evidence.

\begin{equation}
    R_j(0) = r_j^{\text{init}} 
    %\quad \text{where } r_j^{\text{init}} \in [0.6, 0.95]
\end{equation}

At runtime, the reliability score is updated as:
\begin{equation}
R_j(t) = \beta \, R_j(t-1) + (1-\beta) \mathbf{1}_{match}
\end{equation}

\begin{equation}
\mathbf{1}_{match} 
\,
\begin{cases}
1, & T_j^{\text{obs}} \leq (1+\delta)\,\hat{T}_j \\
0, & \text{otherwise}
\end{cases}
\label{1match}
\end{equation}

To capture temporal dynamics in network behavior, a memory factor ($\beta \in [0.8, 0.99]$) is introduced to govern the decay of past observations in the reliability score. Since communication latency can fluctuate due to transient congestion, jitter, or short-lived interference, relying on individual observations may lead to disproportionate reliability shifts. $\beta$ mitigates this by smoothing fluctuations and emphasizing long-term trends. A lower value yields a highly reactive system, which heavily penalizes edge servers after minor spikes, while a higher $\beta$ provides stability, ensuring that trust is earned or lost gradually. This property is particularly important in time-varying edge environments, where temporary degradation should not immediately disqualify a server~\cite{josang2007survey}. Moreover, by maintaining a running history of performance, the memory factor promotes fairness~\cite{xiong2004peertrust}: servers affected by temporary background load are not permanently penalized, but can regain reliability if their performance improves. In equation~\ref{1match}, $\delta$ represents a tolerance on the accuracy of the prediction model.

When making server selection decisions, relying solely on predicted latency risks favoring fast servers but increasing handovers, while focusing exclusively on reliability may lead to long execution times. Therefore, to balance these competing objectives, the system model defines a composite score $S_j$ defined as:
\begin{equation}
    S_j = \alpha \cdot \frac{\hat{T}_j}{T_{\max}} + (1 - \alpha) \cdot (1 - R_j),
    \label{eq:tradeoff}
\end{equation}
where:
\begin{itemize}
    \item $\hat{T}_j$: Predicted latency for $\phi{_j}$,
    \item $T_{\max}$: Maximum predicted latency in \( \mathbf{E} \),
    \item $R_j$: Current reliability score of $\phi{_j}$,
    \item $\alpha \in [0,1]$: Weighting factor that controls the tradeoff between delay and reliability.
\end{itemize}

\begin{algorithm}[t]
\caption{MO-HAN Server Selection}
\label{alg:offloading_reliability}
\begin{algorithmic}[1]
%\REQUIRE $D_E[1\dots n]$, $D_{\text{proc}}^{\text{edge}}[1\dots n]$, $\delta_{\max}$, %$R_j$, $\alpha$, $\theta_{\text{handover}}$

\STATE Compute $\hat{T}_{1\dots n}$
\STATE $C \gets \{\hat{T}_j \mid j=1,\dots,n\}$
\STATE $T_{\max} \gets \max_{j} \hat{T}_j$

\FORALL{$j \in C$}
    \STATE $S_j \gets \alpha \cdot \dfrac{\hat{T}_j}{T_{\max}} + (1 - \alpha) \cdot (1 - R_j)$
\ENDFOR

\STATE $\phi_{j_{\text{best}}} \gets \arg\min_j S_j$

\IF{$\phi_{j_{\text{best}}} = \phi_{j_{\text{prev}}}$}
    \STATE $\phi_j^* \gets \phi_{j_{\text{prev}}}$ \COMMENT{No change needed}
\ELSIF{$S_{j_{\text{prev}}} - S_{j_{\text{best}}} < \theta_{\text{handover}}$}
    \STATE $\phi_j^* \gets \phi_{j_{\text{prev}}}$ \COMMENT{Maintain current connection}
\ELSE
    \STATE $\phi_j^* \gets \phi_{j_{\text{best}}}$ \COMMENT{Handover}
\ENDIF

\RETURN $\phi_j^*$ 

\end{algorithmic}
\end{algorithm}

%%%%%%%%%%%%%%%%%%%%%%%%%%

%%%%%%%%%%%%%%%%%%%%%%%%%%%%%%

To avoid frequent handovers due to small score differences, we introduce a switching-aware threshold $\theta_{\text{handover}}$. The system compares the current server’s composite score $S_{j_{\text{prev}}}$ to the lowest available score $S_{j_{\text{best}}}$, and a handover is made only if:

\begin{equation}
S_{j_{\text{prev}}} - S_{j_{\text{best}}} \geq \theta_{\text{handover}}
    \label{eq:theta_tradeoff}
\end{equation}

This factor introduces hysteresis by stabilizing decisions to ensure that handover transitions only occur when the performance gain is meaningful. 

The model with prediction-guided server selection, adaptive reliability, and switching awareness with hysteresis is implemented in the Moderate-Handover (MO-HAN) algorithm, presented in Algorithm~\ref{alg:offloading_reliability}.

\section{Experimental Setup and Results}
\label{sec:setup}

\subsection{Testbed Overview}

\begin{figure}[htbp]
  \centering
  \includegraphics[width=0.85\linewidth]{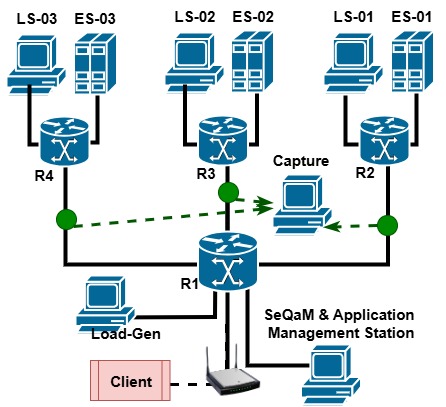} % Update with your actual image path
  \caption{Experimental setup  with dynamic load and real-time monitoring.}
  \label{fig:topology}
\end{figure}

We deployed a custom edge computing testbed to evaluate the proposed methodology under dynamic and realistic network conditions. The architecture of the experimental setup is illustrated in Fig.~\ref{fig:topology}.

\subsection{Components and Functionality}

\begin{itemize}
    \item \textbf{Client Device (ESP32-WROOM):} Sends synthetic data at random intervals to emulate a real-time communication between the embedded device and the edge.
    
    \item \textbf{Load Generator (Load-Gen):} 
    Generates stochastic TCP/UDP background traffic with dynamic load variations, directing flows to load servers (LS) to simulate real-world network congestion.
    
    \item \textbf{Routers (R1–R4):} Provide connectivity between the client, servers, and monitoring station. Routing decisions steer tasks towards candidate edge servers, while background flows are distributed across the topology.
    
    \item \textbf{Capture Node:} A passive monitoring station equipped with \texttt{tcpdump}/\texttt{TShark}, continuously logs packet-level traces to extract metrics such as packet arrival rate (pps), throughput, and link utilization.
    \item \textbf{Edge Servers (ES):} Dedicated to processing tasks. These servers execute computation-intensive workloads such as inference or data analytics.
    \item \textbf{Load Servers (LS):} Act as traffic endpoints for the Load-Gen, thus introduce dynamic interference traffic, producing variable congestion patterns across the network.
     \item \textbf{Client Application Management System:} A Service Quality Manager for Edge Computing (SeQaM)~\cite{burbano2024end} functions as an edge monitoring and management framework that collects real-time data from capture stations and transmits it to the Client Application Management System (CAMS). CAMS then analyzes key metrics and makes server selection decisions for the client.
\end{itemize}

\subsection{Data Collection}
The data collection was designed to capture both the variability of network conditions and the responsiveness of the system under realistic time-varying traffic. Two processes were executed in parallel during the experiments. First, the Load Generator produced random TCP/UDP traffic streams directed toward the Load Servers. This ensured that link utilization and packet arrival rates exhibited stochastic variations similar to those observed in real deployments. Second, the Client Device continuously transmitted data frames whose sizes were randomly drawn from a uniform distribution in the range of 400–600~KB. Each frame was offloaded to one of the candidate edge servers for processing. We measured the observed end-to-end latency for every transmission, which is defined as the elapsed time between the client’s transmission and the reception of the processed result. In parallel, the Capture Node recorded packet-level traces using \texttt{tcpdump}. 

The collection experiment was conducted repeatedly under varying load conditions until a dataset of more than 5000 samples was collected. Specifically, every record contains the client frame size, the measured packet arrival rate, the observed link utilization, and the corresponding end-to-end latency experienced by the frame. This dataset was subsequently used to train and validate the proposed latency prediction models as well as to evaluate the MO-HAN algorithm in comparison to existing methods. By combining synthetic load generation with real-time passive monitoring, the collected data reflects a wide range of network conditions, thereby ensuring robustness and generalization of the trained model.

\subsection{Operational Flow}

The process operates in the following stages:

\begin{enumerate}
    \item When the client has a frame to process, it first sends the frame size to the monitoring station (CAMS).
   
     \item The monitoring station, through SeQaM, collects the latest \texttt{Arrival\_rate} and \texttt{Utilization} metrics from the capture nodes. 
     \item These metrics are then forwarded to CAMS, which uses a pre-trained model to predict the expected latency $\hat{T}_j$ for each candidate $\theta_j$.
  
    \item The server selection algorithm is executed.
    
    \item The client receives the selected target $\theta_j^*$ and forwards the request accordingly.
    
    \item The observed end-to-end latency is logged and used for further model evaluation and reliability adjustment.
\end{enumerate}

\subsection{Server Selection Algorithms}

To evaluate the effectiveness of our proposed methodology, we benchmark it against several representative server selection strategies commonly considered. The algorithms are implemented as follows:

\begin{enumerate}
    \item \textbf{Nearest [NR-HAN]:} Serves as a ground baseline in which the device remains permanently associated with the nearest edge server based on physical proximity. This approach reflects the simplest deployment strategy often adopted in practice, as it minimizes control overhead and avoids handovers. 
    
    \item \textbf{Round Robin [RR-HAN]:} Implements a cyclic allocation policy where requests from the device are distributed sequentially across the available edge servers. This strategy promotes a coarse form of load balancing by ensuring that no server is systematically neglected.

    \item \textbf{Lowest Latency-Server Handover [LL-HAN]:} The device selects the server with the minimum latency offered. This policy represents a purely opportunistic, greedy strategy that pursues the higher-performing edge server to process the device request. 
  
\end{enumerate}

By contrasting these baselines with our approach, we highlight the limitations of static and purely reactive policies in dynamic edge environments.

\subsection{Results}

\begin{figure}[htbp]
\centering
\includegraphics[width=0.5\textwidth]{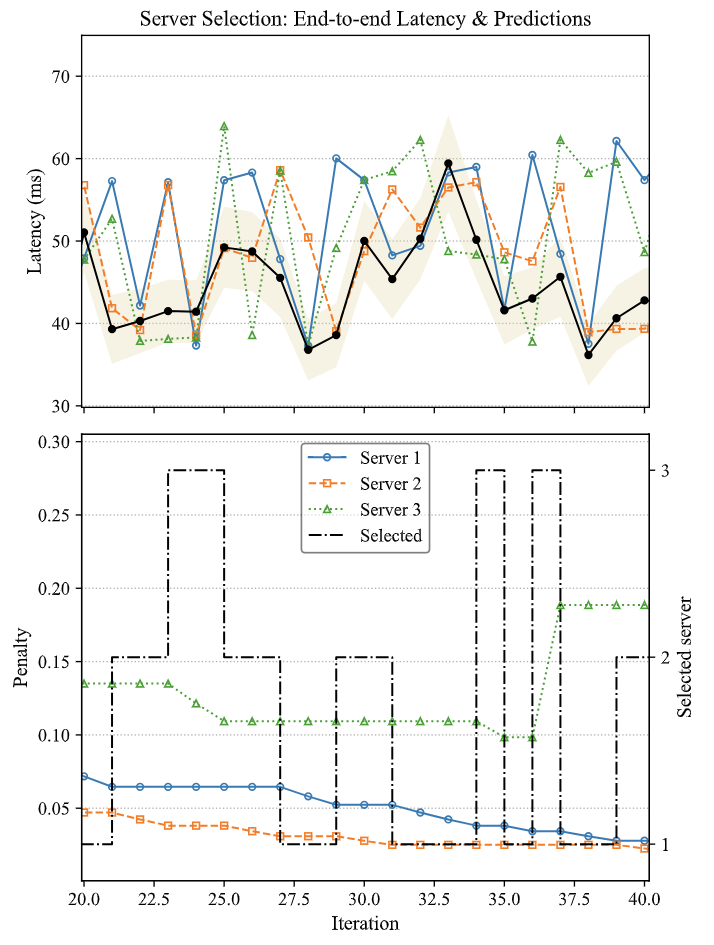}
\caption{Server selection by implementing the proposed MO-HAN algorithm.}
\label{fig:scenario2}
\end{figure}

A sensitivity study over the collected data demonstrates that the selector’s knobs in MO-HAN play competing roles: increasing $\alpha$ improves tail latency by prioritizing predicted delay but raises the number of handovers; larger $\theta$ suppresses handovers via hysteresis at the cost of worse $P95$ latency under bursts; and $\delta$ moderates trust updates, with moderate values ($0.15$--$0.25$) balance jitter tolerance and penalize persistent under-prediction..
These objectives are thus inherently conflicting, yielding a Pareto set rather than a single optimum. Fig.~\ref{fig:scenario2} shows the obtained results from applying MO-HAN algorithm with $\alpha = 0.5$, $\beta = 0.9$, $\delta = 0.2$, and $\theta = 0.05$. This near-frontier setting serves as an example that balances prediction-driven agility with moderate reliability tolerance and light hysteresis, yielding low $P95$ latency with acceptable handover. The top panel compares predicted latencies $\hat{T}_j$ with the observed end-to-end latency and the tolerance of the selected server, while the bottom panel tracks penalties and server choices. The algorithm adapts to servers offering significantly lower $\hat{T}_j$ even when penalized, but issues new penalties when observed latency deviates beyond $\delta$, triggering switches if reliability degrades. The memory factor $\beta$ gradually decays penalties, avoiding permanent exclusion after transient spikes, thereby balancing short-term latency gains with long-term reliability.

\begin{figure}[htbp]
\centering
\includegraphics[width=0.48\textwidth]{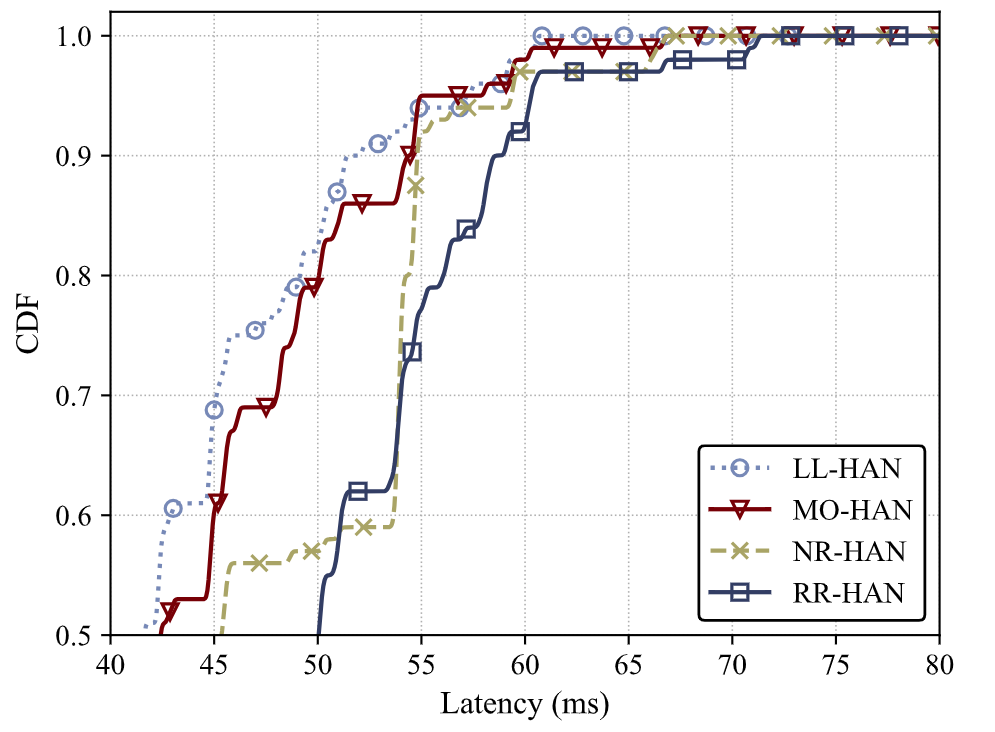}
\caption{Cumulative distributed function of end-to-end latency with the different server selection algorithms.}
\label{fig:scenario1}
\end{figure}

For the fair comparison of the algorithms, the exact same dataset is used to perform the server selection decisions. Both end-to-end latency and handover rate (e.g., percentage of handovers performed by an algorithm) are the two main metrics employed for comparison. Fig.~\ref{fig:scenario1} compares the cumulative distribution of end-to-end latency achieved by the evaluated server selection algorithms. The results show that MO-HAN consistently outperforms NR-HAN and RR-HAN, achieving lower latencies across the entire distribution. MO-HAN performs more effectively by identifying and maintaining connections to responsive servers, whereas NR-HAN and RR-HAN frequently suffer from suboptimal choices.

Compared with the opportunistic LL-HAN, MO-HAN attains better latency for 95\% of the requests while reducing the number of handovers by almost half (37.4 vs. 68.7). This demonstrates that its adaptive reliability and switch-awareness with hysteresis design not only improves latency but also suppresses unnecessary switches, which otherwise introduce transient performance spikes when a new server is engaged. Importantly, our experiments do not account for re-authentication costs, meaning LL-HAN’s frequent switching would likely incur even higher delays in real deployments. Table~\ref{tab:comparison} summarizes the key statistical metrics derived from the evaluation of each server selection algorithm. Overall, MO-HAN strikes a superior balance between low latency and decision stability, making it better suited for resource-constrained embedded applications.


\begin{table}[htbp]
\centering
\scriptsize
\begin{threeparttable}
\caption{Comparison of execution times and handover rate across algorithms}
\label{tab:comparison}
\begin{tabularx}{\columnwidth}{lXXXXX}
\toprule
Metric & u & MO-HAN & RR-HAN & LL-HAN & NR-HAN \\
\midrule
$\mu (T^{obs})$     & ms    & 44.1 & 49.5 & 43.3 & 47.1 \\
$\tilde{T}^{obs}$   & ms    & 42.4 & 50.1 & 41.6 & 45.4 \\
$P_{95}(T^{obs})$   & ms    & 55.0 & 60.2 & 57.1 & 59.4 \\
$HR$                & \%    & 37.4 & 98.9 & 68.7 & 0.0 \\
\bottomrule
\end{tabularx}
\begin{tablenotes}
\footnotesize
\item \textbf{Note: }$\mu$: mean, $\tilde{T}$: median, $P_{95}$: 95th percentile, $HR$: handover rate.
\end{tablenotes}
\end{threeparttable}

\end{table}

\section{Conclusions} \label{sec:conclusions}
This work introduced MO-HAN, a Moderate Handover algorithm for dynamic server selection in edge applications. By combining latency prediction with adaptive reliability scoring and hysteresis-based switching, MO-HAN balances short-term responsiveness with long-term stability. Experimental evaluation confirms that the proposed algorithm suppresses unnecessary handovers, and maintains lower end-to-end latency compared to static and reactive policies. Its lightweight design makes it suitable for deployment on embedded IoT devices with constrained resources. Future research will extend this framework toward multi-objective optimization, incorporating energy efficiency and task criticality, as well as scaling evaluations to federated edge scenarios.

\section*{Acknowledgment}
This work was conducted as part of the project EMULATE, funded by the European Union and the German Federal Ministry of Economics and Climate Action under research grant 13IPC012.

\bibliographystyle{IEEEtran}
\bibliography{Latency}

\end{document}